\documentclass[aps,prl,twocolumn,longbibliography]{revtex4-2}

\usepackage{amssymb}
\usepackage{amsbsy}
\usepackage{amsmath}
\usepackage{graphicx}
\usepackage{graphics}
\usepackage{setspace}
\usepackage{array}
\usepackage{color}
\usepackage{fontenc}
\usepackage{textcomp}
\usepackage{bm}
\usepackage{float}
\usepackage[bookmarks=false,linkcolor=blue,urlcolor=blue,colorlinks,citecolor=blue]{hyperref}

\DeclareMathOperator{\sgn}{sgn}

\newcommand{\vex}[1]{\bm{\mathrm{#1}}}

\newcommand{\blue}[1]{{\color{blue}{#1}}}

\newcommand{\bsub}{\begin{subequations}}
\newcommand{\esub}{\end{subequations}}

\graphicspath{{../Figures/}}

\begin{document}
\title{Dynamical Generation of Higher-order Spin-Orbit Couplings, Topology and Persistent Spin Texture in Light-Irradiated Altermagnets}
\author{Sayed Ali Akbar Ghorashi}\email{sayedaliakbar.ghorashi@stonybrook.edu}
    \affiliation{Department of Physics and Astronomy, Stony Brook University, Stony Brook, New York 11974, USA}
\author{Qiang Li}
    \affiliation{Department of Physics and Astronomy, Stony Brook University, Stony Brook, New York 11974, USA}
    \affiliation{Condensed Matter Physics and Materials Science Division, Brookhaven National Laboratory, Upton, New York 11973-5000, USA}

\date{\today}

\newcommand{\be}{\begin{equation}}
\newcommand{\ee}{\end{equation}}
\newcommand{\bea}{\begin{eqnarray}}
\newcommand{\eea}{\end{eqnarray}}
\newcommand{\h}{\hspace{0.30 cm}}
\newcommand{\vs}{\vspace{0.30 cm}}
\newcommand{\n}{\nonumber}

\begin{abstract}
Altermagnets have been identified as the third category of magnetic materials, exhibiting momentum-dependent spin splitting characterized by even powers of momentum. In this study, we show that when subjected to elliptically polarized light, these materials serve as an exemplary framework for the dynamic generation of topological bands featuring higher-order spin-orbit coupling (SOC). Notably, while the generated Zeeman field remains invariant to the particular altermagnetic ordering, the induced higher-order SOCs are related to the magnitude and symmetry of the altermagnetic order. Specifically, we show that an altermagnet exhibiting $\vex{k}^n$-spin splitting can generate spin-orbit couplings up to $\vex{k}^{n-1}$. In the limit of circularly polarized light, the only correction is $k^{n-1}$, with all lower-order contributions being nullified. Interestingly, light-induced SOCs significantly impact the low-energy band topology, where their Chern numbers change by $\Delta C =\pm 1,2,3$ for $d,g,f$-wave altermagnets. 
Finally, we find a critical field in which a persistent spin texture is realized, a highly desirable state with predicted infinite spin lifetime. Our work showcases light as a powerful, controllable tool for engineering complex and exciting phenomena in altermagnets.   
\end{abstract}

\maketitle

\blue{\emph{Introduction}}.---
Altermagnetism is an emerging magnetic phase that is characterized by an alternating non-relativistic momentum-dependent spin splitting while maintaining vanishing net magnetization \cite{altermagnet1,altermagnet2,altermagnetismreview, song2025altermagnets}. On a microscopic level, they originate from magnetic sublattices connected through rotation rather than translation or inversion, which is typical of conventional collinear antiferromagnets. Dictated by symmetry, altermagnets (ALMs) can exhibit d-, g-, or i-wave harmonics with spin splittings of $k^2$, $k^4$ and $k^6$ order, respectively. while these spin splittings are enforced by symmetry in the absence of spin-orbit coupling (SOC), $\lambda$, (i.e., ideal altermagnetic limit), they remain intact in the presence of $\lambda < J$ where $J$ is the altermagnetic order. In fact, SOC has been shown to be crucial for some of the exciting properties of ALMs such as transport and topology \cite{fernandes2023topological,ghorashiALMnonlinear,ghorashiALMTSC,PhysRevB.110.024425,PhysRevLett.134.096703}.\\ 
The close connection between crystalline symmetry and magnetic order in ALMs creates numerous opportunities for engineering and controlling a variety of properties and responses. Recently, efforts have been made to manipulate these properties in equilibrium using electric fields, strain, or chemical design \cite{mazin2023inducedmonolayeraltermagnetismmnpsse3,strainALM1,strainALM2,fender2025altermagnetism,wei2024crystal,zhou2024crystaldesignaltermagnetism,zhou2025manipulation,wang2024electric}.

\begin{figure}[ht!]
    \centering
    \includegraphics[width=1\linewidth]{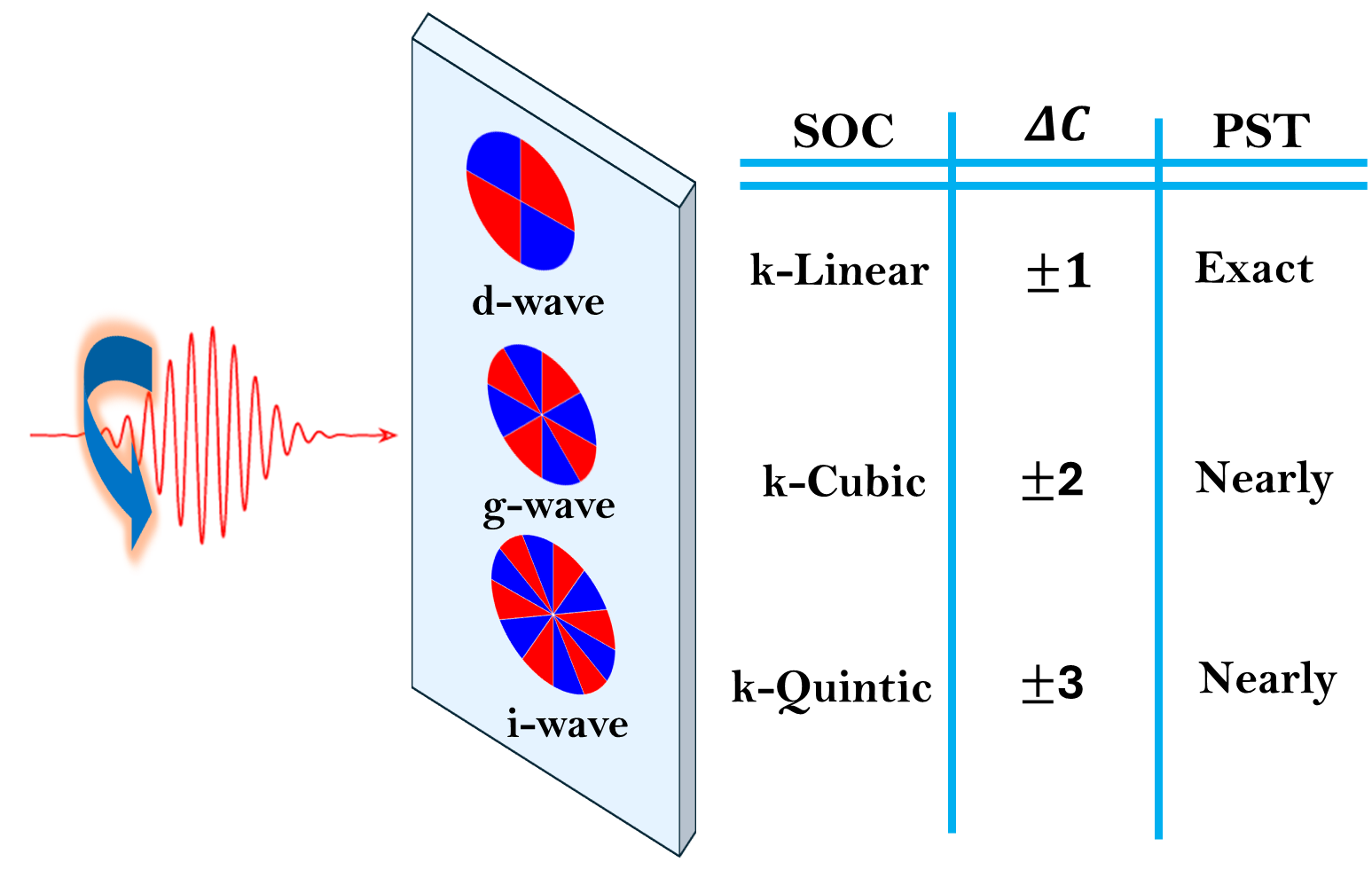}
    \caption{\textbf{Summary of results}. Light generates linear, cubic, quintic-k SOC in $d, g, i$-wave ALMs. The generated SOCs change Chern numbers of the low-energy bands by $\Delta C= \pm 1,2,3$ for $d, g, i$-wave ALMs. At a critical field an exact (nearly) PST emerges in $d$-wave ($g,i$-wave) ALMs.}
    \label{fig:adpic}
\end{figure}
In this work, we demonstrate the non-equilibrium engineering of ALMs using the Floquet formalism. We systematically investigate the effect of elliptically polarized light (EPL) on planar \(d\), \(g\), and \(i\)-wave ALMs, revealing three key findings. First, we show that irradiated ALMs offer a tunable platform for generating anisotropic and higher-order odd in-\(k\) SOCs. In particular, for circularly polarized light (CPL), the dominant correction takes the form \(k^{n-1}\), where \(n\) is an even integer representing the power of altermagnetic momentum dependence, with all lower-order contributions being suppressed. Second, we demonstrate that ALMs provide a promising route for achieving a tunable persistent spin texture (PST), which has been predicted to support extraordinarily long spin lifetimes for carriers, an attractive feature for spintronics and quantum information applications \cite{PSTreview,PSTbernevig,PST1,pst2,pst3,pst4}. Finally, we discuss how light can be harnessed to engineer the topology and band geometry in ALMs. In particular, we show a systematic change in valley Chern bands where the higher-order SOCs result in higher changes in Chern bands. Fig.~\ref{fig:adpic} presents a summary of our results.

\begin{figure*}[ht!]
    \centering
    \includegraphics[width=1\linewidth]{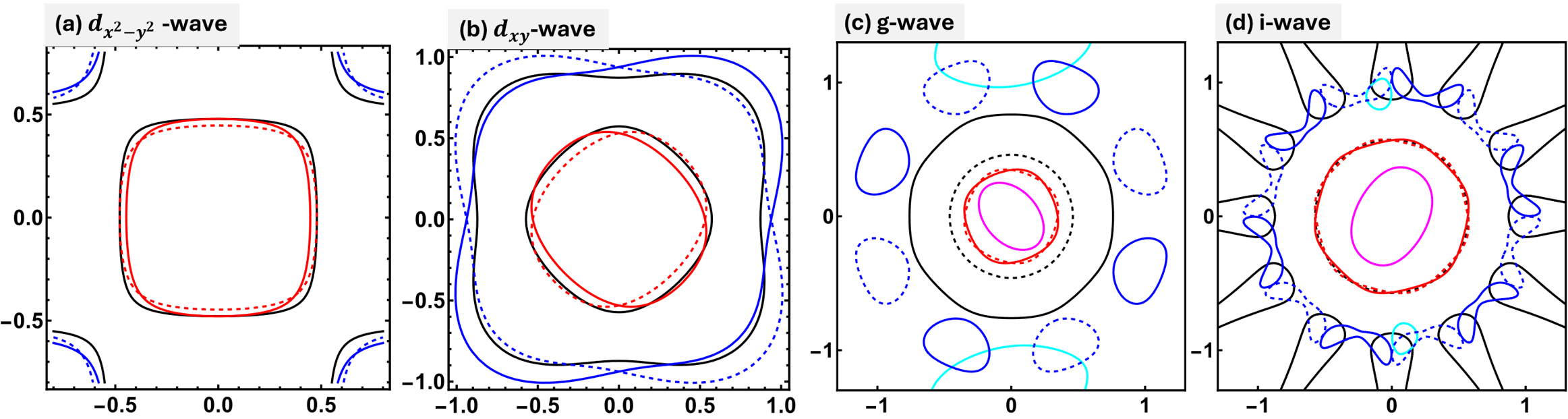}
    \caption{\textbf{Fermi surface of irradiated planar altermagnets}.The Fermi surface of altermagnets with (a) $d_{x^2-y^2}$-wave order for $\mu=0.5$, (b) $d_{x^2-y^2}$-wave order for $\mu=0.5$, (c) $g$-wave order for $\mu=0.35$, (d) $i$-wave order for $\mu=0.5$. (Black solid and dashed), (solid blue and red), (dashed blue and red) and (cyan and magenta) are representing spin up and down FSs for the case with (no), ($\eta=1,A_{x,y}=1.5$), ($\eta=-1,A_{x,y}=1.5$), ($\eta=1,A_x=1.8575, A_y=1$) lights. $t=1, \lambda=0.3, J=1, \omega=5$ is used for all plots.}
    \label{fig:FS}
\end{figure*}
\blue{\emph{Floquet formalism}}.---
We start with shining a general elliptically polarized light,
\begin{align}
\vec{A}(t)=\left(\eta A_x\cos(\omega t), A_y\sin(\omega t)\right)
\end{align}
where $\eta=\pm1$ denotes left and right-handed polarized light. The coupling of light with matter is treated via the Pierls substitution, $H(k)\rightarrow H(k+e\vec{A}(t))$, therefore the full time-periodic Hamiltonian can be written as, $H(k,t)=\sum_nH_ne^{in\omega t}$. The the effective time-independent
Hamiltonian ($\mathcal{O}(1/\omega^2)$) can be obtained as \cite{floq1,floq2,floq3},
\begin{align}
H_{eff}(k)=H_0+\sum_{n\geq1}\frac{\left[H_{+n},H_{-n}\right]}{n\omega}+O(\frac{1}{\omega^2}).
\end{align}

\blue{\emph{$d$-wave altermagnet}}.---We consider two configurations of $d$-wave ALMs: $d_{x^2-y^2}$- and $d_{xy}$-wave. 
A 2D $d_{x^2-y^2}$-wave ALM can be described by the following effective Hamiltonian around the $\Gamma$ point \cite{altermagnet1}, 
\begin{align}
    H_d=t (k_x^2+k_y^2)+\lambda(k_x\sigma^y-k_y\sigma_x)+J_1(\vex{k})\sigma^z,
\end{align}

where $J_1(\vex{k})=J\left(k_y^2-k_x^2\right)$ and $\sigma^i$ are Pauli matrices representing spin space and $\lambda$ and $J$ denote SOC and altermagnetic strength terms. Following the Floquet procedure described above, the light-induced effective Hamiltonian in the presence of elliptically polarized light is obtained as,

\begin{align}\label{effd}
    H^{eff}_{dx2y2}=&\,t (k_x^2+k_y^2) +(\lambda+\lambda')k_x\sigma^y-(\lambda-\lambda')k_y\sigma^x\cr 
    +& [J_1(\vex{k})+J']\sigma^z,
\end{align}
where $\lambda'=2\eta A_xA_yJ\lambda/\omega$, $J'=-\eta A_xA_y\lambda^2/\omega$.

\begin{figure}[t!]
    \centering
    \includegraphics[width=1\linewidth]{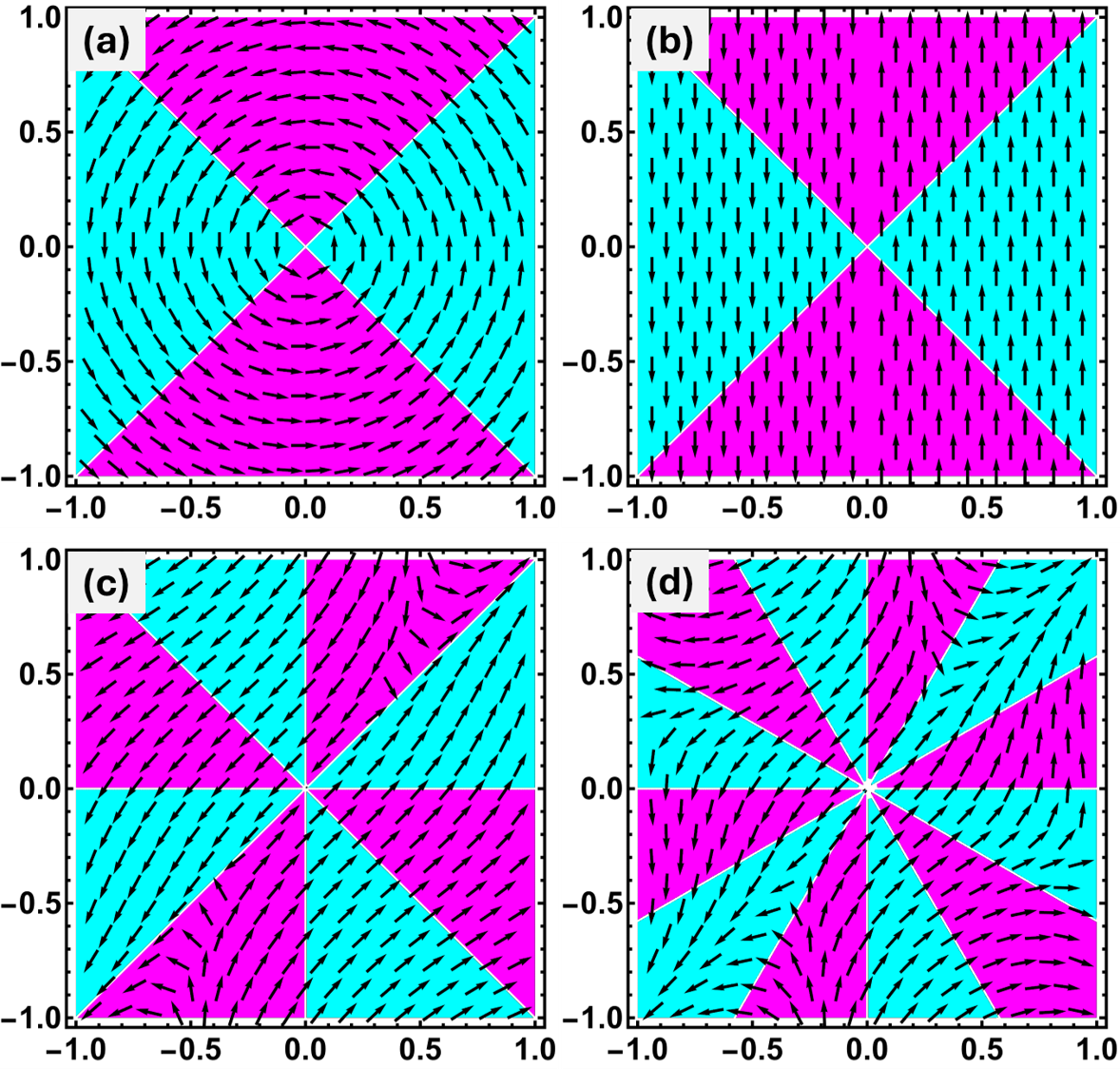}
    \caption{\textbf{Persistent spin texture in altermagnets}. Spin texture in (a) $d_{x^2-y^2}$-wave altermagnet without light (b) $d_{x^2-y^2}$-wave altermagnet at $\bar{A}^c_d$, (c) $g$-wave ALM at $\bar{A}^c_g$, (d) $i$-wave ALM at $\bar{A}^c_i$. Arrows show the in-plane spin-texture and cyan and magenta denote out of the plane spin polarization.}
    \label{fig:PST}
\end{figure}

\begin{figure}[t]
    \centering
    \includegraphics[width=1\linewidth]{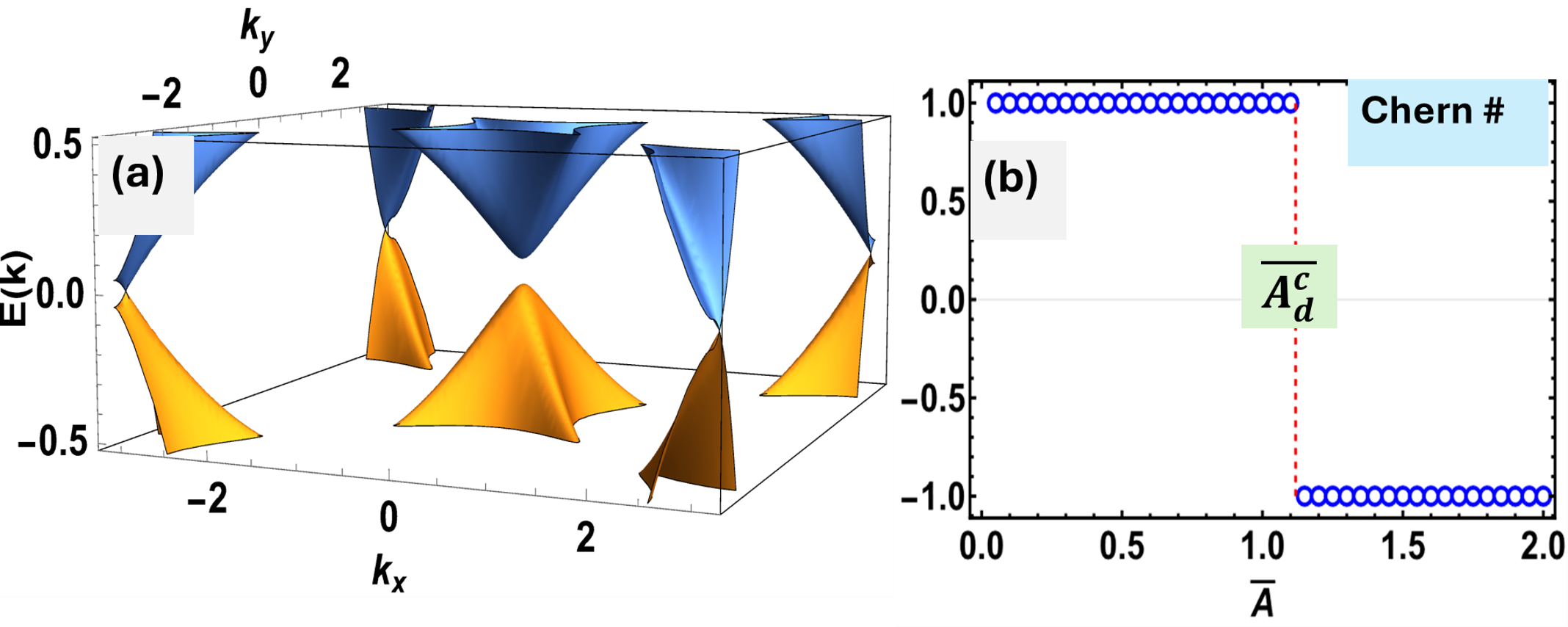}
    \caption{\textbf{Topology and Berry curvature}. (a) Topological phase transition in $d_{x^2-y^2}$-wave ALM at $\bar{A}^c_d=\sqrt{\frac{\omega}{2\alpha \eta J}}$ with $\alpha=1$. (b) Chern number vs $\bar{A}$ with $\alpha=2$. Dashed line shows $\bar{A}^c_d$ with $\alpha=2$. $t=1, \lambda=0.3, J=1, \eta=1, \omega=5$ are used for all plots.}
    \label{fig:topology}
\end{figure}

Interestingly, for $\lambda=0$ (ideal altermagnetic limit) the effect of light vanishes, highlighting the importance of SOC, although $\lambda <<J$, in light-matter coupling in altermagnets. There are two main corrections: (i) a light-induced uniform magnetization, $J'$, that is independent of $J$ and breaks $C_4 T$, as a result opens a gap and induces Chern bands with $C=1/2$ for the low-energy model in \eqref{effd} around the $\Gamma$ point.  Note that there is another $C=1/2$ contribution from the effective model around the $M$ point, so the total Chern number for the lattice model is $C=1$. This term is not unique to ALMs and appears in other systems that possesses Dirac cones, such as graphene and surface states of TI \cite{floq1,floq4}. (ii) A new SOC that induces an imbalance in spin texture along the $x$ and $y$ directions and leads to Fermi surface (FS) anisotropy as can be seen from Fig.~\ref{fig:FS}(a). This term locks the polarization of the light, $\eta$, to spin and vanishes for $J=0$, reflecting its altermagnetic origin. As a result of the spin-light locking, the FS anisotropy can be switched with light. It should be noted that in the altermagnetic limit of $\lambda << J$, $\lambda'>> J'$ and thus (ii) is a more dominant effect. \\
The tunable anisotropy in SOC has a remarkable effect on the underlying spin texture of the ALMs. For $\lambda=\pm |\lambda'|$ the SOC along $k_y$ direction vanishes leading to an in-plane persistent spin texture (PST) where the spin configuration in the momentum space becomes uniform as shown in Fig.\ref{fig:PST}(b). Interestingly, the direction of this PST can be selected at will by controlling $\sgn(\eta J)$. Setting $A_y=\alpha A_x=\alpha \bar{A}$ and $\eta J > 0$, we find $\bar{A}^c_d=\sqrt{\frac{\omega}{2\alpha \eta J}}$ as the critical laser field to achieve the PST. \\
The induced anisotropic SOC also significantly affects the underlying topology of altermagnets. This can be seen from the Berry curvature, 
\begin{align}\label{bc}
    \Omega_{xy}=\frac{-(\lambda-\lambda')^2 \left(J_1(\vex{k})+J'\right)}{2 \left((J(\vex{k})+J')^2+k_x^2 (\lambda +\lambda')^2+k_y^2 (\lambda-\lambda')^2\right)^{3/2}},
\end{align}
 For $\lambda'=\lambda$ exactly where PST appears, the Berry curvature vanishes hinting a topological transition. Indeed, as shown in Fig.~\ref{fig:topology}(a), at $\lambda=\lambda'$ a $d_{x^2-y^2}$-wave ALM undergoes a topological phase transition denoted by the vanishing of the gap in two isolated points away from $k=0$. The gap for $\lambda'>\lambda$ reopens with a reversed sign of the Chern number \ref{fig:topology}(b). For finite $\lambda'$ the valley Chern number deviates from its quantized value of $|1/2|$, nevertheless, the full Chern number in the lattice model remains quantized. Note that in Fig.~\ref{fig:topology}(a), we have used a lattice model, as the low-energy model around the $\Gamma$ point in \eqref{effd} is incapable of capturing the closure of the gap. 
 
 Let us now consider how the physics described above would differ for a $d_{xy}$-wave ALM with $J_2(\vex{k})=J k_xk_y\sigma^z$ spin splitting. The Floquet Hamiltonian reads as,
 \begin{align}\label{effdxy}
   H^{eff}_{dxy}=&\,t (k_x^2+k_y^2) +\lambda(k_x\sigma^y-k_y\sigma^x)-\lambda'(k_x\sigma^x-k_y\sigma^y)\cr 
    +& [J_2(\vex{k})+J']\sigma^z  
 \end{align}
The first difference is that the generated anisotropic SOC has a rotated spin-texture. This is also reflected in the modified FS, Fig.~\ref{fig:FS}(b). Importantly, this leads to the emergence of PST along the $k_x-k_y (k_x+k_y)$ directions instead of $k_x(k_y)$ in the case of $d_{x^2-y^2}$-wave ALM. \\ 
The topological aspects differ even more. First, while the Hamiltonian in \eqref{effdxy} exhibits \( |C| = 1/2 \), the full lattice model of a \( d_{xy} \)-wave ALM is topologically trivial. This occurs because unlike the $d_{x^2-y^2}$-wave case, the contributions from all high-symmetry points cancel out each other. The total Chern number remains zero even after the phase transition point $\lambda=\lambda'$. Similarly to $d_{x^2-y^2}$, the valley quantization is only intact for either $\lambda'<<\lambda$ or $\lambda'>>\lambda$. It is important to note that this key difference between $d_{x^2-y^2}$ and $d_{xy}$ is not limited to Floquet physics discussed here and holds for the general case of $C_4T$ breaking mass terms that can open a gap in $d$-wave ALMs, which seems to have been overlooked in previous literature. Finally, we emphasize that the non-trivial valley band geometry and topology around $\Gamma$ and other high-symmetry points could have important implications for valleytronics \cite{valley1,valley2,valley3,valley4}.

\blue{\emph{g-wave altermagnet}}.---
Now we consider a minimal effective model for $g$-wave altermagnet as \cite{altermagnet1},
\begin{align}
    H_g=t (k_x^2+k_y^2)+\lambda(k_x\sigma^y-k_y\sigma^x)+J k_x k_y(k_x^2-k^2_y)\sigma^z,
\end{align}
and obtain the Floquet effective Hamiltonian as, 
\begin{align}\label{effg}
    H^{eff}_g= H_g + J'\sigma^z +& \lambda' \Bigg(\frac{3}{8}(A_y^2-A_x^2)(k_x\sigma^x-k_y\sigma^y)\cr
    -&\frac{1}{2}\left[
  \begin{array}{cc}
    0 & k_+^3 \\
    k_-^3 & 0 \\
  \end{array}
\right]
\Bigg)
\end{align}
Similar to the case of \(d\)-wave ALMs, two main effects are the induction of a gap and anisotropic SOC. However, unlike \(d\)-wave ALMs, light generates both a cubic SOC term and a linear term.  \\
An interesting feature emerges when considering the limit of CPL, where $A_x = A_y$. In this case, the generated linear SOC term vanishes, and only the cubic term remains, creating a distinct SOC structure. However, for general EPL, both linear and cubic SOC terms can coexist, leading to a more complex interplay between these two components. This is obvious from the light-modified FSs which are clearly distinct in presence of EPL and CPL, Fig.~\ref{fig:FS}(c).\\
Unlike the $d$-wave ALMs, the induced cubic SOC prevents an exact PST, however, because the cubic term is more relevant at higher momenta, a near-low-energy PST can still occur in the presence of EPL. Up to linear in $k$, we obtain the critical laser field $\bar{A}^c_g=\sqrt{\frac{2\sqrt{\omega}}{\sqrt{3\alpha(\alpha^2-1)\eta J}}}$ that PST emerges. As shown in Fig.~\ref{fig:PST}(c), the effect of cubic SOC on PST is negligible. 

Let us now consider how light influences the topology and band geometry of a g-wave ALM. Three major contributions are the uniform Zeeman term and the induced linear ($\lambda^L$) and cubic SOCs ($\lambda^{cubic}$). Although these three terms are not independent, to gain a better understanding of the topology, it is instructive to first consider the effect of each term independently. Setting $\lambda^{linear}=\lambda^{cubic}=0$, in the presence of $J'$ the model in \eqref{effg} exhibits $|C|=1/2$ bands. 
Now, if we ignore the cubic term and only manually turn on the induced linear SOC, similar to $d$-wave ALMs, the system undergoes a phase transition at $\lambda = \lambda^{linear}$, resulting in a reversal of the sign of the Chern bands, i.e., $\Delta C = \pm 1$. 
Note that this is the condition under which the (only approximately when both $\lambda^{linear},\lambda^{cubic}$ are nonzero) PST occurs. As discussed above, this condition cannot be satisfied in the current Floquet setup, as for EPL, both $\lambda^{linear}$ and $\lambda^{cubic}$ always coexist.
On the other hand, we showed that for CPL, the $\lambda^{linear}$ term vanishes, leaving only the cubic contribution. Remarkably, in this case, the Chern number changes by two, $\Delta = \pm 2$, meaning that the bands possess either $|C| = 5/2$ or $|C| = 3/2$. Therefore, controllable light-induced higher-order SOCs have a significant impact on the band geometry and topology by inducing higher wavefunction windings.         

\blue{\emph{i-wave altermagnet}}.---
We repeat similar steps for an $i$-wave altermagnet with following Hamiltonian \cite{altermagnet1},
\begin{align}\label{Hi}
    H_i=t (k_x^2+k_y^2)+&\lambda(k_x\sigma^y-k_y\sigma^x)\cr
    +&J k_x k_y(3k_x^2-k^2_y)(3k_y^2-k^2_x)\sigma^z,
\end{align}
and obtain an effective light-matter Hamiltonian as
\begin{align}\label{effi}
    H^{eff}_i=& H_i + J'\sigma^z \cr
    &+ \lambda' \Bigg(\frac{15}{16}(A_x^2-A_y^2)^2\left[
  \begin{array}{cc}
    0 & k_+ \\
    k_- & 0 \\
  \end{array}
\right]\cr
    &+\frac{15}{4}(A_x^2-A_y^2)\left[
  \begin{array}{cc}
    0 & k_+^3 \\
    k_-^3 & 0 \\
  \end{array}
\right]\cr
&+\frac{3}{2}\left[
  \begin{array}{cc}
    0 & k_+^5 \\
    k_-^5 & 0 \\
  \end{array}
\right]
\Bigg).
\end{align}
Interestingly, in the case of $i$-wave ALMs, light generates SOCs up to the $5^{th}$ order and FSs show an even more anisotropic structure, Fig.~\ref{fig:FS}(d). Furthermore, in the presence of CPL, all subleading SOCs vanish, resulting in a purely quintic SOC. Quintic SOCs are even more rare in materials compared to cubic SOCs. Therefore, irradiated ALMs provide a unique platform to controllably generate and study such higher-order SOCs.\\ 
Like $g$-wave ALMs, as a result of induced higher-order SOCs, only an approximate PST can occur. We obtain a critical $\bar{A}^c_i=\sqrt{2} (\frac{\omega}{15 \alpha (-1 + \alpha^2)^2 \eta J})^{1/6}$ for the emergence of approximate PST up to linear in $k$. As shown in Fig.~\ref{fig:PST}(d) for small momenta in PST is survived, though expectedly,in a lesser way compared to $g$-wave ALMs. 

Investigating the topological character of irradiated $i$-wave ALMs could be even more subtle considering that there are linear, cubic, and quintic SOCs. For the sake of simplicity, we only consider the case of a CPL where only the quintic term exists. 
In the presence of only a Zeeman term, the bands in Eq.\eqref{Hi} exhibit $|C| = 1/2$. Interestingly, by turning on the quintic SOC, the Chern number jumps to $|C| = 5/2$ or $|C| = 7/2$, depending on the relative sign of $J'$ and the quintic SOC, resulting in a total change of $|\Delta C| = 3$. Therefore, we demonstrate that the induced higher-order SOCs systematically change the band geometry and topology by inducing higher windings of the wavefunction at the $\Gamma$-valley in irradiated ALMs.




\blue{\emph{Concluding remarks}}.---we have shown that light induces anisotropic SOCs in ALMs, with higher harmonics of the magnetic order generating higher-order SOCs. Moreover, there is a clear distinction between CPL and EPL for the $g,i$-wave order, where all subleading light-induced SOCs vanish in the CPL case. Importantly, despite having different harmonics, for CPL the light-induced corrections in all ALMs are $\mathcal{O}(\bar{A}^2)$. \\
The PST discovered in this work introduces ALMs as a completely new and rich class of material candidates for the controllable realization of these much-desired phases.\\
Finally, we have shown that ALMs provide a suitable platform for a tunable and systematic enhancement of wavefunction windings and band geometry. A complete study of the topology of lattice models in more general setups will be presented elsewhere.

\emph{Acknowledgments}.---
The work at Stony Brook University was supported by SUNY Research Foundation for Stony Brook University. Qiang Li acknowledges the support by the U.S. Department of Energy, Office of Basic Energy Sciences, Contract No. DE-SC0012704.\\

\emph{Note:} During final stage of this work, we became aware of another work \cite{yarmohammadi2025anisotropiclighttailoredrkkyinteraction} in which a similar effective Hamiltonian is obtained for $d$-wave ALMs. Apart from this, the physics and focus of the two works are very different. 

\bibliography{Floquet_altermagnet}

\end{document}